# Enhancement of Curie temperature in $Ga_{1-x}Mn_xAs/Ga_{1-y}Al_yAs$ ferromagnetic heterostructures by Be modulation doping

T. Wojtowicz[a],

Department of Physics, University of Notre Dame, Notre Dame, IN 46556 and
Institute of Physics, Polish Academy of Sciences, 02-668 Warsaw, Poland

W.L. Lim, X. Liu, M. Dobrowolska, and J. K. Furdyna

Department of Physics, University of Notre Dame, Notre Dame, IN 46556

K. M. Yu and W. Walukiewicz

Electronic Materials Program, Materials Sciences Division,
Lawrence Berkeley National Laboratory, Berkeley, CA 94720

I. Vurgaftman and J. R. Meyer

Code 5613, Naval Research Laboratory, Washington, DC 20375

ABSTRACT

The effect of modulation doping by Be on the ferromagnetic properties of $Ga_{1-x}Mn_xAs$ is investigated in $Ga_{1-x}Mn_xAs/Ga_{1-y}Al_yAs$ heterojunctions and quantum wells. Introducing Be acceptors into the $Ga_{1-y}Al_yAs$ barriers leads to an increase of the Curie temperature $T_C$ of $Ga_{1-x}Mn_xAs$, from 70 K in undoped structures to over 100 K with the modulation doping. This increase is qualitatively consistent with a multi-band mean field theory simulation of carrier-mediated ferromagnetism. An important feature is that the increase of $T_C$ occurs only in those structures where the modulation doping is introduced *after* the deposition of the magnetic layer, but not when the Be-doped layer is grown first. This behavior is expected from the strong sensitivity of Mn interstitial formation to the value of the Fermi energy during growth.

---

[a] Author to whom correspondence should be addressed; electronic mail: wojto@ifpan.edu.pl

The recent emergence of ferromagnetic $III_{1-x}Mn_xV$ semiconductors has led to a number of innovative application concepts [1]. Since "spintronic" devices will be much more attractive if they are non-cryogenic, a primary focus of the research on $Ga_{1-x}Mn_xAs$ and related materials has been on enhancing the Curie temperature $T_C$. It is by now widely accepted that the ferromagnetic interaction is mediated by free holes [2], and mean field theory projects that $T_C$ should scale as the product of the Mn concentration and the hole density. A number of investigations have sought to maximize the density of Mn ions that can be incorporated substitutionally rather than as compensating interstitials [3]. Yet even after annealing, hole concentrations tend to be much smaller than the Mn densities, and the resulting $T_C$ values often fall short of those projected by mean-field theory [4].

Here we study the alternative strategy of leaving the Mn concentration fixed while increasing the hole density through *p*-type modulation doping. The influence of *p*-type modulation doping on the magnetic properties has so far only been reported for δ-doped GaAs:Mn structures [5] grown at high temperature, and never for a low-temperature-grown $III_{1-x}Mn_xV$ alloy. In this paper we show that remote *p*-type doping can substantially increase the Curie temperature of $Ga_{1-x}Mn_xAs$ in heterojunction and quantum-well (QW) geometries. Such structures may find use in gated devices whose magnetism is efficiently controlled by an applied electric field [6]. The total number of Be atoms in the barrier was varied by two methods: changing the Be flux while fixing the width of the doped region ($d_{Be}$) in the barrier, and changing $d_{Be}$ for a constant Be flux. We show that it makes a critical difference whether the Be-doped $Ga_{1-y}Al_yAs$ barrier layer is deposited *before* or *after* the magnetic layer is grown. This provides additional confirmation of our earlier proposal based on Be co-doping experiments on $Ga_{1-x}Mn_xAs$, that $T_C$ is limited by the value of the Fermi energy during growth [7-9].

Ferromagnetic $Ga_{1-x}Mn_xAs/Ga_{1-y}Al_yAs$ heterostructures were grown on semi-insulating (001) GaAs substrates in a Riber 32 R&D molecular beam epitaxy system. Prior to growing the actual structure, a 450 nm GaAs buffer layer was deposited at 590°C, followed by 3 nm of GaAs grown at 210°C. We then deposited either a $Ga_{1-x}Mn_xAs/Ga_{1-y}Al_yAs$ heterojunction or a $Ga_{1-y}Al_yAs/Ga_{1-x}Mn_xAs/Ga_{1-y}Al_yAs$ QW at the same low temperature of 210°C. Three distinct series of samples were grown. The first two series were heterojunctions, consisting of 5.6 nm of $Ga_{1-x}Mn_xAs$ followed by a 13.5 nm $Ga_{0.76}Al_{0.24}As$ barrier. The barriers were modulation-doped with Be at a distance of 1 monolayer away from the $Ga_{1-x}Mn_xAs$. In Series 1 (with $x = 0.066$), $d_{Be}$ was kept constant at 13.2 nm, and the Be concentration was varied by changing temperature of the Be cell ($T_{Be} = 0$, 1020, 1040 and 1050°C). To provide an estimate of the a hole concentration in the doped regions, Hall measurements were performed on a thick ($d = 98$nm) $Ga_{0.76}Al_{0.24}As$:Be layer grown at $T_{Be}= 1050$°C, which gave $p = 3.1 \times 10^{20}$ cm$^{-3}$. In Series 2 (with $x = 0.062$), $T_{Be}$ was kept at a constant 1040°C, and three different doping-layer thicknesses were employed ($d_{Be} = 0$, 5.3, and 13.2 nm).

In addition to the heterojunctions we also grew three $Ga_{0.76}Al_{0.24}As/Ga_{1-x}Mn_xAs/Ga_{0.76}Al_{0.24}As$ quantum well structures (Series 3), with Be doping in one of the two barriers (either *before* or *after* the QW). The width of the QW was 5.6 nm, $x = 0.062$, the total thickness of each barrier was 13.5 nm, and $d_{Be} = 13.2$ nm. Resistivity, Hall effect, and SQUID measurements were used for electrical and magnetic characterization of the samples, and for determining $T_C$.

Temperature-dependent zero-field resistivities $\rho(T)$ for samples from Series 1 and 2 are presented in Figs. 1(a) and (b), respectively. Data for a 67-nm-thick $Ga_{1-x}Mn_xAs$ epilayer with the same Mn concentration as Series 1 ($x = 0.066$) are also shown as the open points in Fig. 1(a).

All of the samples show a clear resistivity maximum at a temperature $T_\rho$, which may be taken as a convenient estimate of $T_C$ [9]. While all of the undoped samples (including several grown under the same conditions, but not shown in the figure) have very similar values of $T_\rho$ (around 75 K), the doped structures all have higher $T_\rho$, ranging from 95 K ($T_{Be}$ = 1020°C) to 110 K ($T_{Be}$ = 1050°C). To further test the enhancement of $T_C$ with increasing Be concentration in the barrier, we studied heterojunctions in which the thickness of the Be-doped region $d_{Be}$ was varied while keeping the Be flux constant (Series 2). Temperature-dependent resistivities for that series are shown in Fig. 1(b). One can see that $T_\rho$ increases from 77K in the undoped structure ($d_{Be}$ = 0) to ≈ 98K for $d_{Be}$ = 5.3 nm, and to 110K for $d_{Be}$ = 13.2 nm.

That the modulation doping increased $T_C$ is further corroborated by anomalous Hall effect (not shown) and direct SQUID magnetization measurements. SQUID results for three of the samples from Series 1 are given in Fig. 2, which shows the temperature dependence of the remanent magnetization $M$, measured in an in-plane magnetic field of 100 G. These data confirm the unambiguous increase of $T_C$ with increasing Be modulation doping. Values of $T_C$, taken to be the temperature at which $M(T)$ drastically changes the slope before vanishing, are indicated in the figure, along with the corresponding values of $T_\rho$. The good agreement between the two lends further support for the assumption that $T_\rho$ represents a good estimate of $T_C$.

As expected from our model [3,7,8], it is essential that the barrier containing the Be atoms be grown *after* the $Ga_{1-x}Mn_xAs$ layer rather than *before*, which may be understood as follows. We showed in earlier experiments on $Ga_{1-x}Mn_xAs$ [3], and also on $Ga_{1-x}Mn_xAs$ co-doped with Be [8,9], that the Fermi energies achievable in this material cannot exceed a certain maximum value $E_{Fmax}$, corresponding to a maximum hole concentration $p_{max}$. This occurs because the relationship between the creation energies for negatively-charged defects (such as the desired substitutional Mn acceptors, $Mn_{Ga}$), and positively-charged defects (such as the unwanted interstitial Mn double donors, $Mn_I$), is controlled by the Fermi energy. When $E_F$ in the $Ga_{1-x}Mn_xAs$ reaches $E_{Fmax}$ due to the increasing free hole density, $Mn_{Ga}$ formation becomes energetically too unfavorable, and a high concentration of compensating $Mn_I$ defects begins to form instead. Previous ion-channeling experiments revealed this type of interstitial Mn creation whenever $p$ increased, due to either a higher Mn concentration [3] or to Be co-doping [8,9]. But a strong increase of the $Mn_I$ concentration at the expense of $Mn_{Ga}$ is also expected whenever $E_F$ increases due to hole transfer from a modulation-doped barrier layer. The $Mn_I$ creation is deleterious to the ferromagnetism for multiple reasons: (1) compensation by the double donors reduces the hole concentration, (2) interstitial Mn is RKKY-inactive (due to negligible $p$-$d$ exchange) [10], and (3) $Mn_I$ forms antiferromagnetic pairs with $Mn_{Ga}$ [3,10], reducing further the density of Mn ions participating in the ferromagnetism. Therefore, any increase of the $Mn_I$ concentration leads to lower $T_C$.

Raising $E_F$ via hole transfer from the Be-doped barrier naturally is thus expected to affect the creation of $Mn_I$ only when the additional holes are already present during the $Ga_{1-x}Mn_xAs$ growth, *i.e.*, when the Be-doped layer is grown first. Thus in the case of $Ga_{1-y}Al_yAs/Ga_{1-x}Mn_xAs/Ga_{1-y}Al_yAs$ QWs one expects an increase of $T_C$ when the doped barrier is grown *after* the $Ga_{1-x}Mn_xAs$ deposition (because then $p$ increases without any influence on the already-incorporated concentration of ferromagnetically-active Mn ions), but a drop of $T_C$ when the Be-doped barrier is deposited *before* the QW (because the larger value of $E_F$ induces in this case a higher $Mn_I$ concentration).

To demonstrate experimentally that the sequencing of the layers plays a crucial role in the $T_C$ enhancement, we grew a series of $Ga_{1-y}Al_yAs/Ga_{1-x}Mn_xAs/Ga_{1-y}Al_yAs$ QW structures (Series

3) in the following order: a structure with Be doping in the first barrier, an undoped structure, and a structure with Be doping in the second barrier. Temperature-dependent resistivities for this series are shown in Fig. 3. As compared to $T_\rho$ ($\approx T_C$) = 87°C for the undoped sample, we find that $T_\rho$ increases as in the previous series for the sample with doping *after* the QW (to 110 K), whereas $T_\rho$ decreases for the sample with modulation doping *before* the QW (to 68 K).

Having established experimentally that the $T_C$ of $Ga_{1-x}Mn_xAs$ can be enhanced by modulation doping (as long as the Be-doped $Ga_{1-y}Al_yAs$ is grown *after the magnetic layer*), it is useful to determine whether a model for carrier-mediated ferromagnetism can reproduce such an increase. We therefore simulated one of the heterojunctions from Series 1 ($T_{Be}$ = 1050°C) using an eight-band effective bond-orbital-method calculation, that self-consistently includes magnetic interactions via mean field theory, as well as charge-transfer-induced electrostatic band bending via Poisson's equation [11]. The hole density in the GaMnAs layer was fit to the experimental finding of $T_C$ = 70 K in the absence of Be doping. For the Be-doped structure this calculation yields $T_C$ = 78 K and $T_C$ = 85 K for two $Ga_{0.934}Mn_{0.066}As/Ga_{0.76}Al_{0.24}As$ valence band offset values: 0.127 eV (corresponding to a vanishing offset between GaAs and unmagnetized $Ga_{1-x}Mn_xAs$) and 0.25 eV, respectively. Note that even with the larger offset, the calculated Curie temperature is substantially smaller than the experimental finding of $T_C$ = 103 K (Fig. 2). Thus, while our calculations do predict an increase of $T_C$ associated with the enhanced hole concentration due to the modulation doping, the agreement is only qualitative. One possible explanation is that the substitutional Be incorporation into the $Ga_{1-y}Al_yAs$ barrier is in fact *enhanced* in the vicinity of the $Ga_{1-x}Mn_xAs$. The reduced Fermi energy in the barrier during its growth (because many of the holes are "siphoned off" into the already-existing QW) should therefore induce an increase of the substitutional $Be_{Ga}$ acceptor concentration as compared to that measured in the doped calibration epilayer (and used in the calculations).

In summary, we have demonstrated the possibility of increasing the Curie temperature of $Ga_{1-x}Mn_xAs$ through modulation doping with Be acceptors. We further showed that in this process the sequence of *first* growing the $Ga_{1-x}Mn_xAs$ layer, and *then* the modulation-doped barrier is critical. The fact that $T_c$ decreases rather than increases in structures doped before the growth of the $Ga_{1-x}Mn_xAs$ provides strong support for the model in which the incorporation of magnetically-active Mn in $Ga_{1-x}Mn_xAs$ alloys, which requires the occupation of substitutional rather than interstitial sites, is governed by electronic considerations [3,7-9].

This work was supported by the DARPA SpinS Program under ONR grant N00014-00-1-0951, by the Director, Office of Science, Office of Basic Energy Sciences, Division of Materials Sciences and Engineering, of the U.S. Department of Energy under Contract No. DE-AC03-76SF00098; and by NSF Grant DMR00-72897.

FIGURE CAPTIONS

Fig. 1 Temperature-dependent zero-field resistivities $\rho(T)$ for $Ga_{1-x}Mn_xAs/Ga_{0.76}Al_{0.24}As$ heterojunctions remotely doped with Be acceptors: (a) Series 1, with $x = 0.066$, $d_{Be}=13.2$ nm, and various $T_{Be}$; (b) Series 2, with $x = 0.062$, $T_{Be} = 1040°C$ and various $d_{Be}$. Sample parameters and peak resistivity values ($T_\rho \approx T_c$) are indicated. Also shown as the open points in (a) are data for a $Ga_{0.94}Mn_{0.66}As$ epilayer with no GaAlAs barrier.

Fig. 2 Temperature dependence of magnetization measured with an in-plane magnetic field of 100 Gs for three $Ga_{1-x}Mn_xAs/Ga_{0.76}Al_{0.24}As$ heterojunctions from Series 1 grown with different Be-cell temperatures $T_{Be}$, indicated in the figure. The values of $T_C$ and corresponding $T_\rho$ are also given.

Fig. 3 Temperature dependence of zero-field resistivity $\rho(T)$ of three $Ga_{0.76}Al_{0.24}As/Ga_{1-x}Mn_xAs/Ga_{0.76}Al_{0.24}As$ quantum well structures from Series 3. Be acceptors were introduced either into the first barrier (that grown *before* the ferromagnetic QW), or into the second barrier; or the sample was undoped, as marked in the figure.

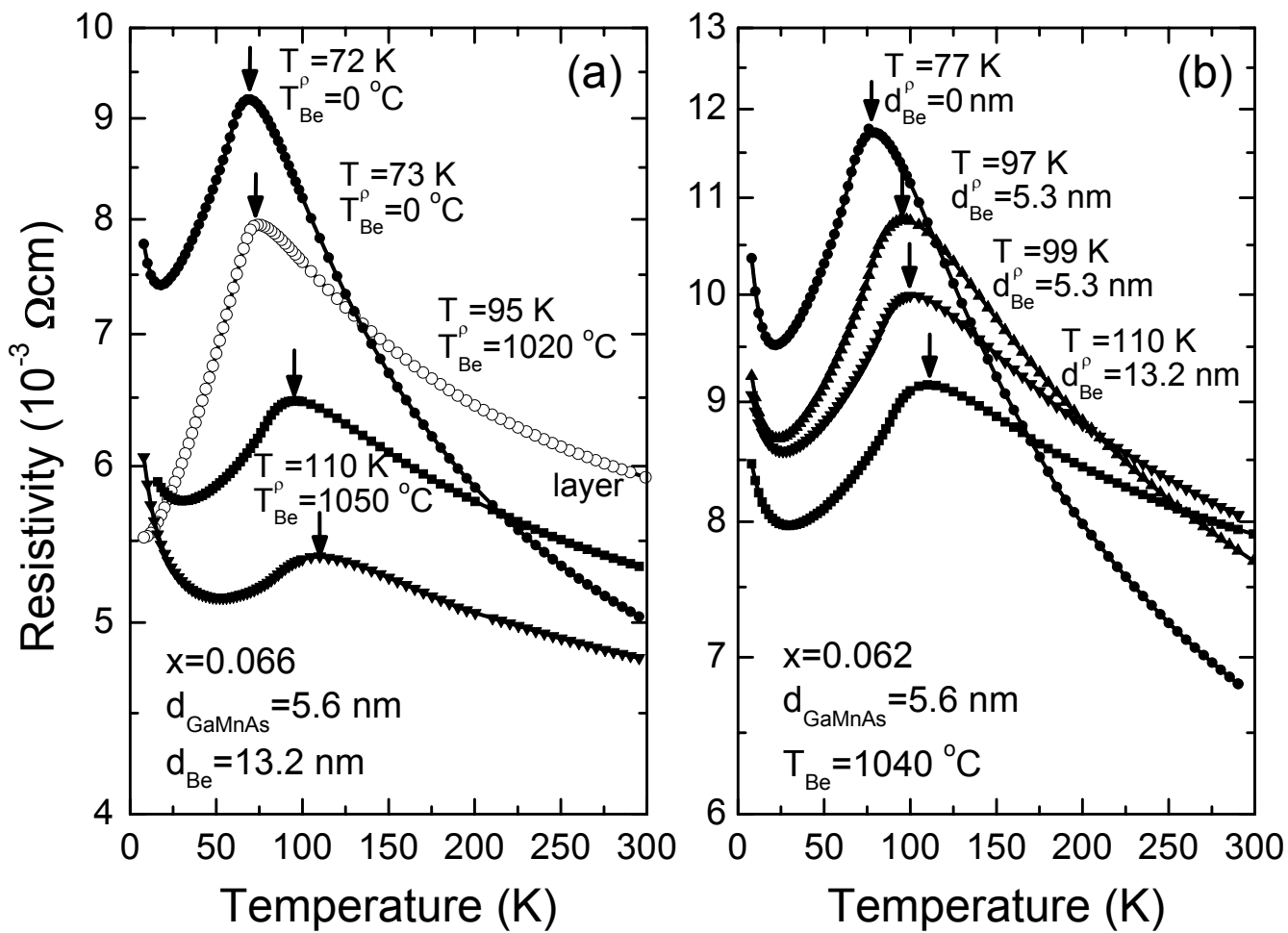

Fig. 1

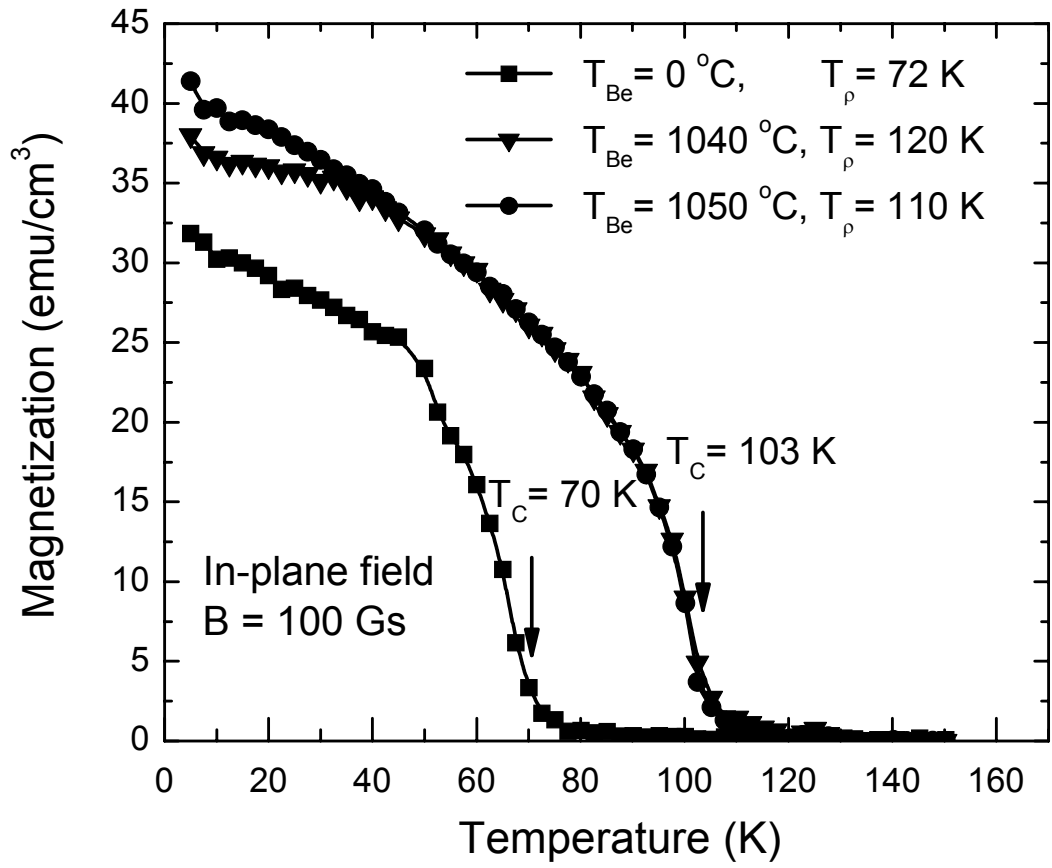

Fig. 2

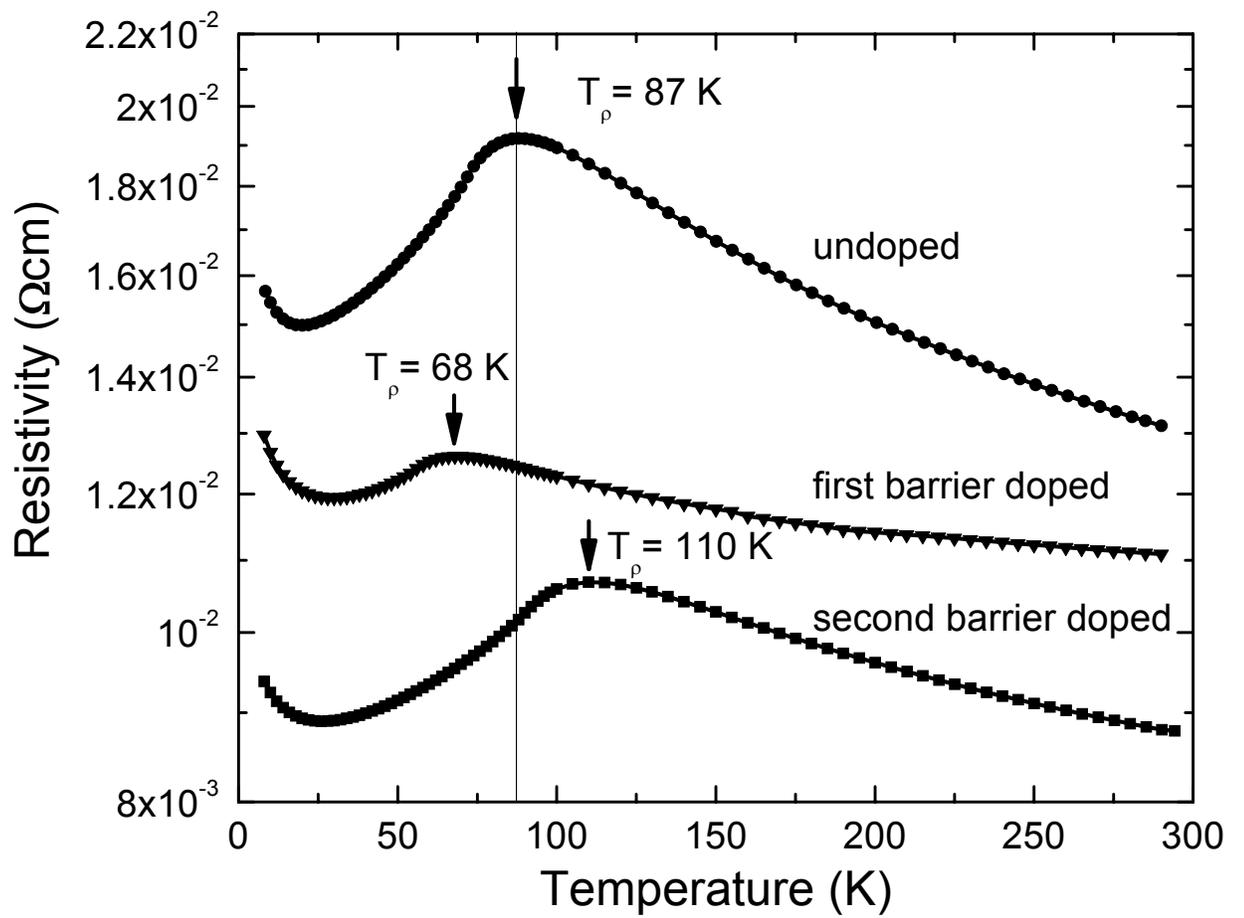

Fig. 3